\providecommand{\LyX}{L\kern-.1667em\lower.25em\hbox{Y}\kern-.125emX\@}

\documentclass[acus]{JAC2000}   
\usepackage{graphics} 
\makeatother

\begin{document}

\title{The RTA Betatron-Node Experiment: \\
Limiting Cumulative BBU Growth In A Linear Periodic System}

\author{S. Lidia, Lawrence Berkeley National Laboratory\\
T. Houck, G. Westenskow, Lawrence Livermore National Laboratory}

\maketitle
\begin{abstract}
The successful operation of a Two-Beam accelerator based on extended relativistic
klystrons hinges upon decreasing the cumulative dipole BBU growth from an exponential
to a more manageable linear growth rate. We describe the theoretical scheme
to achieve this, and a new experiment to test this concept. The experiment utilizes
a 1-MeV, 600-Amp, 200-ns electron beam and a short beamline of periodically-spaced
rf dipole-mode pillbox cavities and solenoid magnets for transport. Descriptions
of the beamline are presented, followed by theoretical studies of the beam transport
and dipole-mode growth.
\end{abstract}

\section{INTRODUCTION}

A Lawrence Livermore National Laboratory (LLNL) and Lawrence Berkeley National
Laboratory (LBNL) collaboration is studying the application of induction accelerator
technology to the generation of microwave power. We refer to this scheme of
power generation as the Relativistic Klystron Two-Beam Accelerator (RK-TBA)
{[}1{]}. This scheme is considered a TBA approach as the extraction of microwave
power is distributed along a drive beam parallel to the high-energy rf linear
accelerator. The RK designation indicates that the power is generated by the
interaction of the relativistic modulated drive beam with resonant structures
similar to those used in a conventional klystron.

The primary advantage of TBA concepts is that the conversion of drive beam power
to microwave power can be highly efficient (\( > \)90\%). This efficiency is
realized by distributing the power extraction over an extended length. The interest
in RK-TBA's is that induction accelerators are efficient at producing very high
power electron beams. Present induction accelerators operate at currents of
several kilo-amperes and accelerate the beam to 10's of MeV for beam power of
100's GW {[}2{]}. The induction accelerator can realize improved efficiency
at converting wall plug power into beam power by replacing the standard electromagnet
solenoids for beam transport with permanent magnets. Even higher efficiency
can be attained if the induction cells are used as high-voltage step up transformers
driven by a relative low voltage (\( \sim  \)20 kV) pulsed power system. Present
designs of a RK-TBA predict efficiency of about 40\% in conversion of wall plug
power into beam power {[}3{]}, or a total wall plug to microwave power efficiency
of about 36\%. The main section of an RK where the microwave power is generated
is comprised of many repeating modules as illustrated in Figure 1. Within each
module, the induction cells replace the energy extracted from the electron beam
by the microwave output structure. The efficiency of this process --- extraction
and reacceleration --- is nearly 100\%. Not shown in Figure 1 are the beam generation
and modulation sections and the final beam dump. Fixed energy losses in those
processes have to be included in calculating the total beam energy to microwave
conversion efficiency. Thus, it is imperative that the RK have many of the efficient
extraction and reacceleration cycles to reduce the relative value of fixed losses
with respect to the total energy transferred to the beam.

Several proof-of-concept experiments have been performed to demonstrate the
viability of the RK-TBA concept. These experiments have shown the generation
of collider-scale drive beam in induction linacs, production of high-quality,
high-power microwaves from standing- and traveling-wave structures driven by
induction accelerator beams, and multiple reacceleration and extraction cycles
{[}4, 5{]}. As will be described below, we are continuing to perform experiments
to study specific physics and technology issues while constructing a prototype
relativistic klystron.

\section{BEAM DYNAMICS ISSUES }

The ultimate efficiency of a RK is determined by the induction beam dynamics
\- i.e. the number of extraction structures that the beam can transit. We have
identified three critical areas of beam dynamics that must be understood. The
first involves maintaining the longitudinal modulation of the beam or \char`\"{}rf
bucket\char`\"{} structure. In the drifts between output structures, space charge
forces will cause the beam to lengthen in phase space, i.e., \char`\"{}debunch\char`\"{}.
If this effect is not corrected, the rf current (Fourier component of the beam
at the modulation frequency) will decrease resulting in a decrease in the microwave
power that can be extracted. Inductively detuning the output structures, similar
to the penultimate cavity in conventional klystrons, can counter the space charge
forces. The requirement for long-term longitudinal stability is reestablishing
the initial longitudinal charge distribution at the end of a synchrotron period.
Computer simulations have shown that with proper detuning, the rf current can
be maintained over the 150 output structures envisioned for a full scale RK-TBA
{[}6{]}.

The other issues involve transverse instabilities. The beam will excite dipole
modes in the induction cell accelerating gaps as well as in the resonant output
structures. The induction cell accelerating gaps can be severely damped with
rf absorbers for all resonant modes since the applied voltage pulse is quasi-static
compared to the resonant frequencies. In addition, the natural energy spread
over the rf bucket contributes to phase mixed, or Landau, damping. The combination
of rf absorbers and energy spread is expected to maintain the transverse instability
due to the dipole modes in the accelerating gaps at acceptable levels. 

The resonant output structures present a more difficult transverse instability
issue. The fundamental mode must couple sufficiently with the beam to extract
the required energy. Various techniques exist to damp higher order modes in
both output and accelerating structures. However, the permanent magnet focusing
system envisioned for an RK-TBA allows the application of a new technique that
we refer to as the \emph{Betatron Node} Scheme. 

Transverse beam instability theory is well developed and the exponential growth
predicted is supported by experiment. However, the standard theoretical approach
assumes that the discrete cavities interacting with the beam are closely spaced
compared to the betatron wavelength due to the focusing system. Our design for
an RK-TBA system requires strong focusing to maintain the required beam radius
and a constant average energy over each extraction/reacceleration cycle. This
combination leads to spacing between output structures of one betatron wavelength
and the basic assumption of the standard theoretical approach does not hold. 

An alternative approach to studying the transverse instability uses transfer
matrices {[}7{]}. Assuming a monoenergetic beam and a thin cavity, Equations
(\ref{Momentum change}) through (\ref{BNS Transformation}) indicate the salient
parts of this theory. Equation (\ref{Momentum change}) represents the transverse
momentum change an electron receives passing through the cavity. \emph{R} is
an integral operator that accounts for the part of the beam that has already
passed through the cavity. The first matrix on the RHS of Equation (\ref{General Transformation })
is then the transfer matrix for the beam going through the cavity. For a sufficiently
thin cavity, the transverse position does not change. Only the momentum is affected.
The second matrix represents the betatron motion of the electrons between cells
where \( \theta  \) is the phase advance. Thus, Equation (\ref{General Transformation })
advances the position and momentum of electrons from the exit of on cavity to
the exit of the following cavity. By repeatedly multiplying the two transfer
matrices, the position and momentum at the exit of any cavity can be related
to the initial conditions. For the situation where \( \theta  \) is constant
for all sections and \( \theta \ll 1 \), the series of matrix multiplications
can be shown to yield the same expected exponential growth as the more standard
approach.
\begin{equation}
\label{Momentum change}
\Delta p_{\bot }=R\cdot x
\end{equation}
 
\begin{eqnarray}
 & \left[ \begin{array}{c}
x\\
p_{\bot }
\end{array}\right] _{n+1_{exit}}= & \nonumber \\
 & \left[ \begin{array}{cc}
1 & 0\\
R & 1
\end{array}\right] \left[ \begin{array}{cc}
\cos \theta  & \frac{\sin \theta }{\omega }\\
\omega \sin \theta  & \cos \theta 
\end{array}\right] \left[ \begin{array}{c}
x\\
p_{\bot }
\end{array}\right] _{n_{exit}} & \label{General Transformation } 
\end{eqnarray}

For the case where \( \theta =2\pi  \) (or any integral multiple of \( \pi  \)),
the matrix multiplication is greatly simplified. The betatron motion returns
the electrons to the original transverse position and momentum (oppositely directed
for odd multiples of \( \pi  \)). The multiplication involves only the matrix
describing the effect of the cavity, and, as shown in Equation (\ref{BNS Transformation}),
this leads to a linear growth in the transverse instability
\begin{eqnarray}
 & \left[ \begin{array}{c}
x\\
p_{\bot }
\end{array}\right] _{n+1_{exit}}= & \nonumber \\
= & \left[ \begin{array}{cc}
1 & 0\\
R & 1
\end{array}\right] ^{n}\left[ \begin{array}{c}
x\\
p_{\bot }
\end{array}\right] _{(n=1)_{exit}} & \nonumber \\
= & \left[ \begin{array}{cc}
1 & 0\\
nR & 1
\end{array}\right] \left[ \begin{array}{c}
x\\
p_{\bot }
\end{array}\right] _{(n=1)_{exit}}\label{BNS Transformation} 
\end{eqnarray}

There are many non-ideal factors in a realistic accelerator including cavities
of finite thickness and variation in phase advance due to energy and/or focusing
errors. Parameter studies through computer simulations indicate that the transverse
instability is significantly reduced for systems with reasonable variations
in parameters. We intend to experimentally test the validity and robustness
of the Betatron Node Scheme.

\section{BETATRON NODE SCHEME EXPERIMENT }

\begin{figure}
{\par\centering \resizebox*{1\columnwidth}{!}{\includegraphics{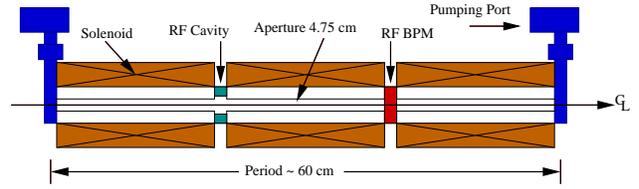}} \par}

\caption{\label{min beamline}Minimum beamline configuration.}
\end{figure}

The basic elements involved in a test of the Betatron Node Scheme are: a set
of devices that generate a localized transverse impedance, a tunable focusing
and transport system, and diagnostics to measure the BBU mode signal on the
beam as a function of time and distance along the beamline. A schematic for
a possible beamline is shown in Figure \ref{min beamline}.

The localized impedances are generated in simple pillbox cavities, tuned so
that the \( TM_{110} \) mode frequency matches the modulation of the beam;
a series of solenoid magnets provide tunable focusing; and rf BPMs placed between
cavities provide a means of collecting the dipole mode signal carried by the
beam. We have built several sections of this beamline, using off-the-shelf components
wherever possible. Each section is one betatron wavelength long and is comprised
of one pillbox cavity, a pumping port, a diagnostic, and three solenoids. The
rf cavities have a simple pillbox design, with a dielectric insert (Alumina
99.5\%, \( \epsilon \approx 9 \)) to adjust the mode frequency. The dipole
mode resonates at \( \sim  \)5.2 GHz, with a wall-loaded Q-value of \( \sim  \)100
and a normalized transverse impedance \( \left[ \frac{Z_{\perp }}{Q}\right] \sim  \)6.5\( \Omega  \).

\begin{figure}
{\par\centering \resizebox*{1\columnwidth}{!}{\includegraphics{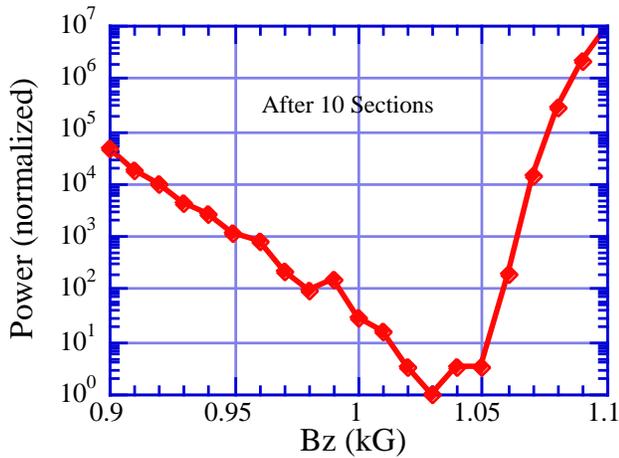}} \par}

\caption{\label{power vs. field}Dipole mode power vs. solenoidal field (phase advance).}
\end{figure}

\begin{figure}
{\par\centering \resizebox*{1\columnwidth}{!}{\includegraphics{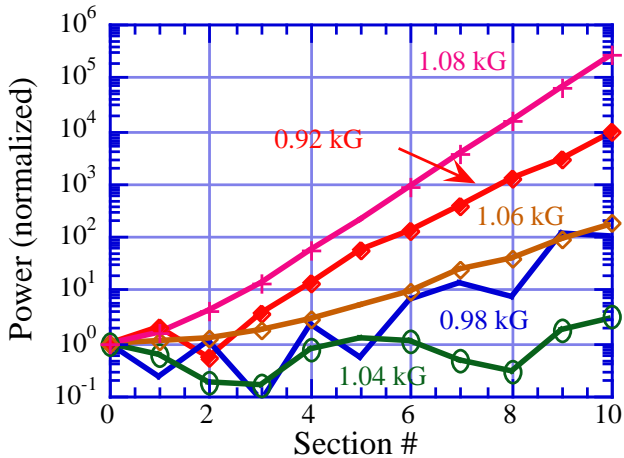}} \par}

\caption{\label{power vs. period}Dipole mode power vs. length for varying solenoid
fields, displaying linear and exponential growth regimes.}
\end{figure}

Computer simulations of the increase in power measured by the rf diagnostics
at the dipole mode frequency are shown in Figures \ref{power vs. field} and
\ref{power vs. period}.
 
 Variations of \( \pm  \)10\% in the solenoidal field (betatron phase advance)
from the optimum should produce several orders of magnitude increase in measured
mode power after only a few sections. The graphs indicate the maximum power
expected during the main body of the beam (\char`\"{}flat top\char`\"{}). The
temporal power variation during the pulse (not shown) is predicted to have different
characteristics between under- and over-focused scenarios.

\section{SUMMARY }

The long-term goal of the RTA Facility is to build a prototype relativistic
klystron that has all the major components required for a RK suitable for collider
applications. The prototype would serve as a test bed for examining physics,
engineering, and cost issues. The first major component, the 1-MeV, 600-A, induction
electron gun, of the prototype has been completed and commissioned. Before continuing
with the next section of the prototype, we intend to perform a series of beam
dynamics experiments. In particular, we will demonstrate the effectiveness of
the Betatron Node Scheme. We are also continuing to study and optimize collider
designs based on the RK-TBA scheme.

\section{ACKNOWLEDGMENTS }

We thank Swapan Chattopadhyay, George Caporaso, Kem Robinson, and Simon Yu for
their support and guidance. Dave Vanecek and Wayne Greenway provided invaluable
mechanical engineering and technical services. John Corlett and Bob Rimmer designed
the rf diagnostics. This work was performed under the auspices of the U.S. Department
of Energy by University of California Lawrence Berkeley Livermore National Laboratory
under contract No AC03-76SF00098 and Lawrence Livermore National Laboratory
under contract No. W-7405-Eng-48.

\section{REFERENCES }

{\par\raggedright 1. Sessler, A.M. and Yu, S.S., \emph{Phys. Rev. Lett.} \textbf{54},
889 (1987). \par}

{\par\raggedright 2. Burns, M.J., et al., \char`\"{}DARHT Accelerator Update
And Plans For Initial Operation\char`\"{}, in \emph{Proceedings of IEEE 1999
Part. Accel. Conf.}, NY, 1999, pp. 617-621. \par}

{\par\raggedright 3. Houck, T.L. (ed.), et al., ``Appendix A: A RF Power Source
Upgrade to the NLC Based on the Relativistic Klystron Two-Beam Accelerator Concept'',
SLAC-474, Stanford University (1996).\par}

{\par\raggedright 4. Westenskow, G.A. and Houck, T.L., \emph{IEEE Trans. Plasma
Sci.} \textbf{22}, pp. 424-436 (1994). \par}

{\par\raggedright 5. G.A. Westenskow and T.L. Houck, ``Results of the Reacceleration
Experiment: Experimental Study of the Relativistic Klystron Two-Beam Accelerator'',
in \emph{Proceedings of the 10th Intl. Conference on High Power Particle Beams},
San Diego, CA (1994). \par}

{\par\raggedright 6. Giordano, G., et. al., ``Beam Dynamics Issues in an Extended
Relativistic Klystron'', \emph{Proceedings of IEEE 1995 Part. Accel. Conf.},
Dallas, TX, 1995, p. 740-742.\par}

{\par\raggedright 7. Neil, V.K., Hall, L.S., and Cooper, R.K., \emph{Part. Accel.}
\textbf{9}, pp. 213-222 (1979).\par}

\end{document}